\documentclass[onecolumn,aps,amsmath,amssymb]{revtex4}

\usepackage{graphicx}
\usepackage{epsf, times}

\newcommand{\dys}{Dy$_2$Ti$_2$O$_7$}
\newcommand{\holm}{Ho$_2$Ti$_2$O$_7$}
\newcommand{\hso}{Ho$_2$Sn$_2$O$_7$}

\begin{document}

\title{Spin Ice, Fractionalization and Topological Order}

\author{C. Castelnovo$^1$, R. Moessner$^2$, and S. L. Sondhi$^3$}

\affiliation{
$^1$ SEPnet and Hubbard Theory Consortium, Department of Physics, Royal Holloway University of London, Egham TW20 0EX, UK\\
$^2$ Max Planck Institute for the Physics of Complex Systems, 01187 Dresden, Germany\\
$^3$ Department of Physics, Princeton University, Princeton, NJ 08544, USA}


\begin{abstract}
The spin ice compounds {\dys} and {\holm} are highly unusual magnets which epitomize a set of concepts of great interest in modern condensed matter physics: their low-energy physics exhibits an
emergent gauge field and their excitations are magnetic monopoles which arise from the fractionalization of the microscopic magnetic spin degrees of freedom. In this review, we provide 
an elementary introduction to these concepts and we survey the thermodynamics, statics and 
dynamics---in and out of equilibrium---of spin ice from these vantage points. Along the way, we touch on topics such as emergent Coulomb plasmas, observable ``Dirac strings'', and irrational charges.
We close with the outlook for these unique materials.
\end{abstract}

\maketitle
%
%

\section{\label{sec: intro}
Introduction
        }
Spin Ice~\cite{Bramwell2001} is a remarkably simple system in some ways---as a first approximation it is simply a classical Ising antiferromagnet. However this simplicity is deceptive. The antiferromagnetism is not directly apparent in the spin variables, it is highly frustrated due to the topology of the lattice and, importantly, it arises from long ranged dipolar forces~\cite{Siddarthan1999}. It turns out that these features combine to
yield a host of properties that put spin ice at the intersection of two particularly interesting streams
of ideas---one of much current interest in quantum condensed matter physics and one of greater antiquity---which make it a much more illuminating system than might have been guessed {\it a priori}.

The first of these streams is invoked by the keywords ``fractionalization'' and ``topological order''.
Fractionalization~\cite{fractionalisation_review} is the phenomenon wherein the quantum numbers of the low lying excitations of a many-body system are non-integer multiples of those of the constituents, e.g. electrons in a
metallic system, or of the natural excitations, e.g. spin flips in an insulating magnet. Perhaps the most celebrated examples of these in recent years are fractionally charged and fractional statistics excitations in quantum Hall phases~\cite{FQH}, although historically solitons in systems in spatial dimension
$d=1$, such as polyacteylene~\cite{Fpoly}, first brought this phenomenon to prominence in condensed matter physics.
The term topological order~\cite{topo_order} is of more recent provenance, although its roots lie in seminal work
in the early 1970s on spin liquids and lattice gauge theories~\cite{pwa-rvb,lattice_gauge}. In the sense in which we will use it,
the term describes ordering characterized by the emergence of a gauge field, as opposed to the emergence of a local order parameter field in broken symmetry phases.~\cite{foot1}
More narrowly, the term
is reserved for cases where the gauge field is governed by a purely topological action but we will
use it in the more expansive sense which is perhaps better called ``gauge order''. Again, the most
celebrated examples of topologically ordered phases in recent years have arisen in quantum Hall
systems~\cite{topoQH,topo_order,foot2}.
This coincidence is not an accident. While in $d=1$ fractionalization is generically associated with soliton formation, in $d>1$ fractionalization is generically associated with topological order.

The second stream of ideas is centered around the quest for magnetic monopoles, dating from the seminal work of Dirac~\cite{Dirac1931}. As a matter of completeness and elegance, electromagnetism would be enhanced by
the existence of particles carrying magnetic charge and it would probably also simplify the teaching
of the subject to undergraduates. Dirac was interested in the constraints placed on possible magnetic
monopoles by quantum mechanics and discovered his celebrated quantization condition. Since that time
monopoles have been unsuccessfully sought in experiment and successfully found in theory, where they
arise naturally in models that go beyond the standard model. Indeed, the current belief is that  monopoles do exist but are extremely massive and exceedingly rare~\cite{mono_searches}.

In this review we discuss spin ice in the light of these two streams of ideas and show how it
gives rise to an emergent gauge field and to fractionalized excitations. These excitations are
monopoles of the emergent gauge field {\it and} monopoles of the microscopic gauge field that implements the magnetostatics of the problem (in an appropriate sense, consistent with the solenoidal character of the microscopic magnetic field). This discussion will provide an especially transparent realization of
topological order in a classical setting, as well as produce condensed matter analogs of monopoles, {\it
and} present a coherent explanation of a set of elegant experiments on these systems.

We begin with a review of the basic facts of life in spin ice---particularly the surprising innocuousness of the dipolar interaction, move on to the monopoles, emergent gauge field
and their dynamics, and conclude with some thoughts on the trajectory of the field.
Along the way we will discuss some relevant experiments.
We caution readers that this is not a proper review of the field but instead a particular perspective
on it. Fortunately, the reader can consult the excellent historic review by Bramwell and
Gingras~\cite{Bramwell2001} as well as a recent detailed one by Gingras~\cite{Gingras2007} and
a brief overview by Balents~\cite{Balents2010}
for complementary surveys.
%
%

\section{\label{sec: spin ice basics}
Spin Ice Basics
        }
\begin{figure}
\includegraphics[width=0.5\columnwidth]{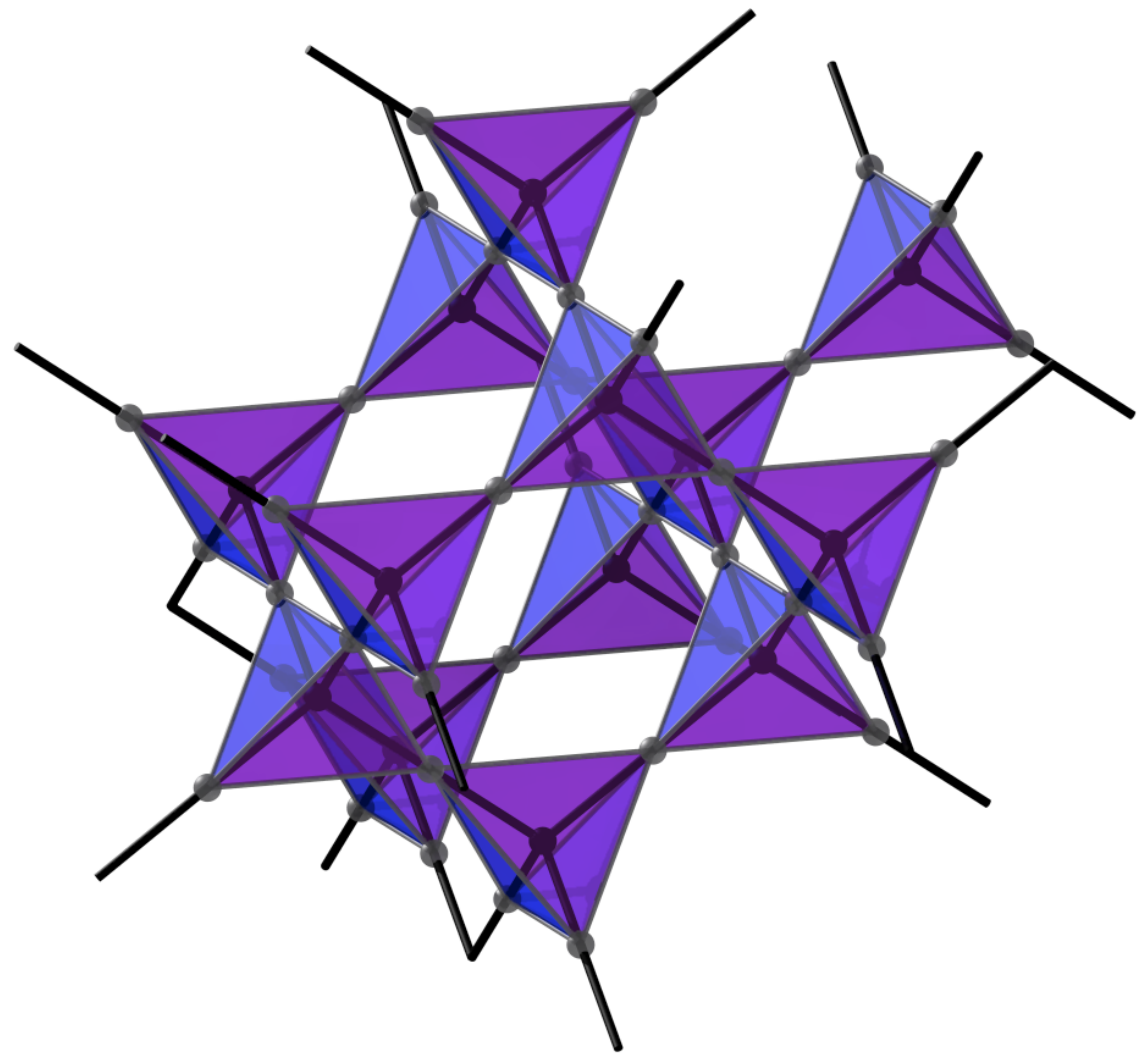}
\caption{
The magnetic moments in spin ice reside on the sites of the pyrochlore lattice, which consists of corner sharing tetrahedra. These sites are at the same time the midpoints of the bonds of the diamond lattice (black) defined by the centres of the tetrahedra. The Ising axes are the local [111] directions, which point along the respective diamond lattice bonds. The bonds of the pyrochlore lattice are in the [110] directions, while a line joining the two midpoints of opposite bonds on the same tetrahedron defines a [100] direction. 
}
\label{fig:pyro}
\end{figure}

The canonical spin ice compounds are {\dys} and {\holm} which host the magnetic ions Dy$^3+$ with
$J=15/2$ and Ho$^3+$ with $J=8$ leading to magnetic moments of the order of $10$ Bohr
magnetons. Both compounds are insulating and the magnetic ions
sit on a sublattice of corner sharing tetrahedra that is commonly referred to as the pyrochlore
lattice (Fig~\ref{fig:pyro}).
Let us begin with the two most salient experimental features of these compounds.
The first, as reported in the original discovery by Harris and Bramwell in neutron scattering~\cite{Harris1997},
is that, despite a ferromagnetic
Curie-Weiss temperature $\approx 2K$, these compounds fail to develop long-range spin order down to a temperature
four times lower. Indeed, they do not order even when cooled further but instead fall out of equilibrium.
\begin{figure}
\includegraphics[width=0.5\columnwidth]{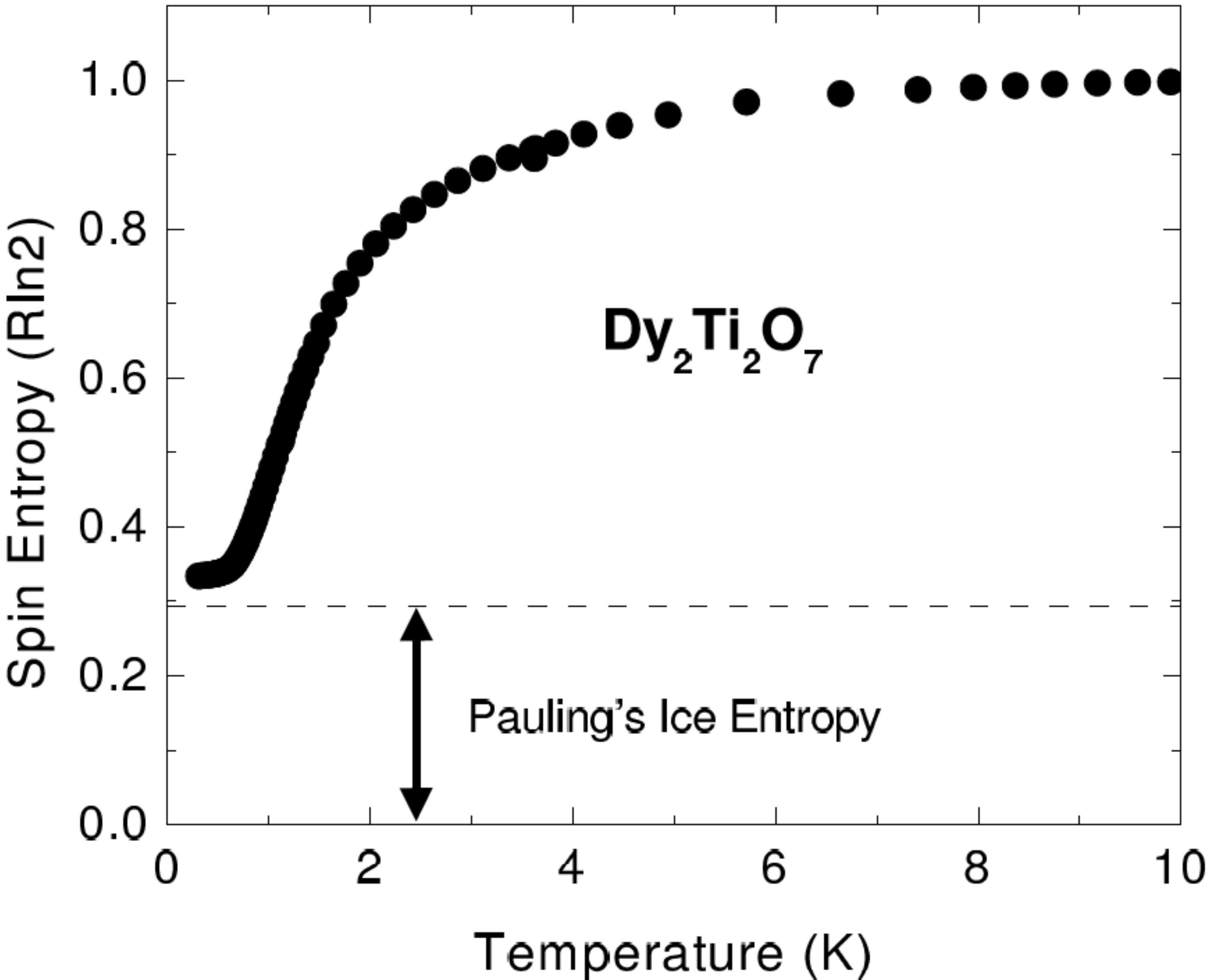}
\caption{
[From Ref.~\cite{Ramirez1999}, Fig.~2b].
Entropy of {\dys} found by integrating the heat capacity divided by temperature from $0.2$~K to $14$~K. The residual entropy at low temperatures is in good agreement with the estimate,
${\cal S}_p/k_B N_s = (1/2) \log (3/2)$, based on Pauling's work on water ice~\cite{Pauling1935}.
}
\label{fig:Ramirez}
\end{figure}
The second, discovered by Ramirez et al~\cite{Ramirez1999} via calorimetry (Fig.~\ref{fig:Ramirez}), is that the spins do not
fully lose their entropy down to the lowest temperatures~\cite{foot3}%
---instead, the spin ice compounds exhibit a
macroscopic entropy per spin, {${\cal S}_0$},  which essentially equals the macroscopic entropy per hydrogen
exhibited
at low temperatures by water ice~\cite{Pauling1935}. These observations are synergistic: the first implies the existence of substantial frustration, the second quantifies it and explains the absence of any phase transition by the failure of the system to hit upon a particularly favorable ordering pattern.
%
%

\subsection{
Disorder, Entropy and Ice Rules
           }
The zeroth order explanation for this behavior is quite simple. The large moments experience a dipolar
interaction which has the right size, about $2K$~\cite{Bramwell2001}. In addition they experience a much larger local [111]
anisotropy, about $300K$, which has the effect of forcing the spins to point either into tetrahedra or
out of them~\cite{Bramwell2001}. The combination of the two terms causes nearest neighbor spins to favor pseudospin antiferromagnetism: on a given tetrahedron an ``in'' spin wishes its neighbors on the same tetrahedron
to point ``out'' and vice versa. Ignoring the longer range of the dipolar interaction, this is really
now an Ising antiferromagnet on the pyrochlore lattice, which has been known to exhibit a macroscopic
entropy at $T=0$ since the ancient work of Anderson~\cite{Anderson1956}.

\begin{figure}
\includegraphics[width=0.5\columnwidth]{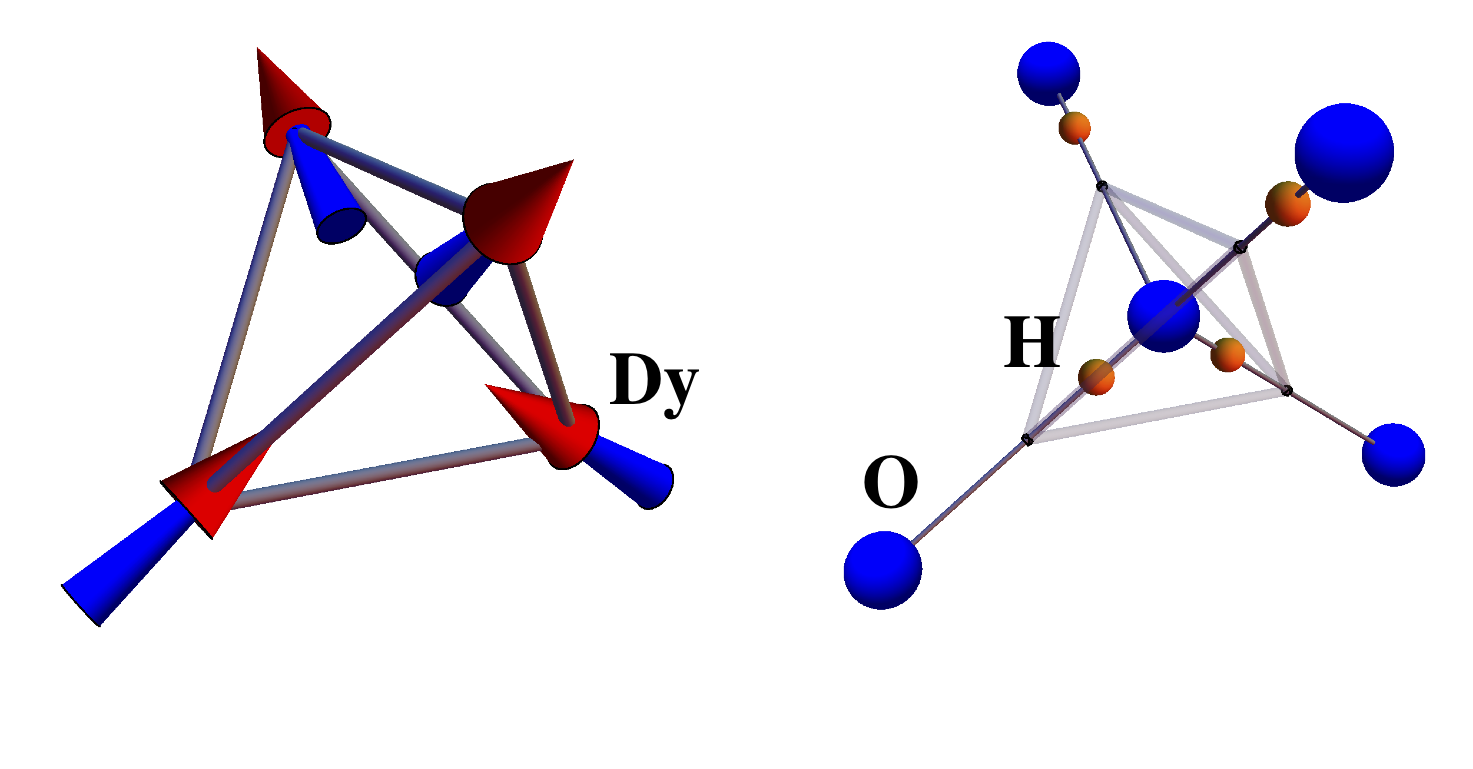}
\caption{
Illustration of the mapping between spin ice and water ice.
A spin that points outwards/inwards indicates an H atom that is displaced away from/towards
the O atom at the center of the tetrahedron.
}
\label{fig:spinwater}
\end{figure}

The entropy itself is accurately estimated by an argument of Pauling's~\cite{Pauling1935}. A given tetrahedron has 6
ground states of the nearest neighbor Ising interaction out of the possible 16 configurations allowed by the crystal field anisotropy. A
lattice with $N_s$ spins and $N_t=N_s/2$ tetrahedra is then estimated to have
\[
2^{N_s} \left({6 \over 16}\right)^{N_t}
\]
ground states via independent application of the tetrahedral constraints. This is equivalent to
a ground state entropy per spin ${\cal S}_p/k_B N_s = (1/2) \log (3/2)$ in fine agreement with
the data in Fig.~\ref{fig:Ramirez}.
This is a good place to note that Pauling's estimate was done in the context of the low
temperature entropy of water ice. The ground states of the two ices are isomorphic upon the
identification of the orientation of the spins with the location of Hydrogen atoms on the bonds
between Oxygens (Fig.~\ref{fig:spinwater}).
In water ice the requirement that exactly two Hydrogens are proximate
to a given Oxygen is the ``ice rule'' that ensures that ice is constructed from water molecules.

As promised, we have explained the macroscopic $T=0$ entropy. The Ising model also explains the
lack of a phase transition at any finite temperature---it has none, and no
signature of long range order develops at any wavevector
all the way down to $T=0$ starting in the high temperature paramagnetic
region.

It would thus appear that we are done, having explained both significant facts about spin ice.
However this is not the case. First, the nearest neighbor Ising antiferromagnet is not so simple
after all---in fact it exhibits a divergent correlation length as $T \rightarrow 0$ which
cuts off {\it algebraic, dipolar} spin correlations which are signatures of an emergent
gauge field~\cite{Huse2003,Moessner2003b,Isakov2004,Hermele2004,Henley2005}. Second, the dipolar interactions among the spins cannot be truncated to nearest
neighbor distance and we need to evaluate their impact on the story we have sketched thus far~\cite{denHertog2000}.
We will begin with the second question and come back to the first one later. We will find eventually
that they lead to a common, comprehensive and yet simple, understanding.
%
%

\subsection{
The Dipolar Puzzle and Its Resolution
           }
To recap: Thus far we have noted that the nearest neighbor dipole-dipole interaction combined
with the [111] easy axis anisotropy yields a pseudospin Ising antiferromagnet with a macroscopic
$T=0$ entropy and also yields the ferromagnetic sign of the Curie-Weiss constant observed in
high temperature measurements~\cite{Bramwell2001}. The ``dipolar puzzle'' is that the
long range part of
the dipolar interaction~\cite{Siddarthan1999}
appears to not significantly change these results. Most importantly, that it does
not lead to low temperature ordering down to $T$ much smaller than the Curie-Weiss constant~\cite{Melko2001,Yavorskii2008}. We will see now that this robustness results from a
remarkable feature of the anisotropy-constrained dipolar interaction on the pyrochlore
lattice---that it differs modestly and only at short range from a ``model dipole'' interaction on
the pyrochlore lattice that has the {\it exact} ground state degeneracy dictated by the ice rules.

There are two very different formulations of this result in the
literature~\cite{Isakov2005,Castelnovo2008} and we will discuss the
second one here as it is essentially pictorial---it leads to the modestly named ``dumbbell
model'' of spin ice (see the Supplementary Information in Ref.~\cite{Castelnovo2008}).
We begin with point dipoles of strength $\mu$ placed on the pyrochlore lattice with their
orientational freedom restricted to the local [111] axes. Now consider replacing each dipole with a dumbbell
consisting of a pair of oppositely charged monopoles of strength $\pm q_m$. These are
placed at distance $d$ in opposite directions away from the pyrochlore lattice site along the
local [111] axis.
At this stage, the
construction is purely mathematical---the monopoles have no reality and we have not changed
the number of degrees of freedom in the system. What we have done is to change the original energy function
written as a sum of $N^2$ dipolar terms 
\begin{equation}
H
=
\frac{\mu_0 \mu^2}{4 \pi} \sum_{i < j}
\left[
  \frac{{\bf S}_i \cdot {\bf S}_j}{ r_{ij}^3}
  -
  \frac{3({\bf S}_i \cdot {\bf r}_{ij})({\bf S}_j \cdot {\bf r}_{ij})}
       {r_{ij}^5}
\right]
\,,
\label{eq:dipH}
\end{equation}
by a monopolar energy function written as a sum of $4N^2$ Coulombic terms
\[
H
=
\sum_{i < j}
\frac{\mu_0}{4 \pi} \frac{q_i q_j}{r_{ij}}
,
\]
where the spins $\bf S_i$ are assumed to point parallel to the local [111] axis, the charges $q_i$ take the values $\pm q_m$, and $\mu_0$ is the vacuum permeability. 

To understand the purpose of the construction consider the interaction between two distant
dumbbells. Up to the constant, immaterial self-energy of each dumbbell, the
Coulombic monopole-monopole
interaction between the constituents of the dumbbells
translates into the original $O(1/r^3)$ dipolar interactions between the spins. Therefore, we have
represented the long ranged part of our original problem faithfully. At not so long distances
there are $O(1/r^5)$ corrections but we can limit their significance at short distances by
two maneuvers. First, we make a clever choice of the separation $d$ between the monopoles---we tune it equal
to the distance $a_d / 2$ to the tetrahedral center (and therefore $q_m = \mu/a_d$).
Second, as this causes monopoles from different dumbbells to overlap, we regularize the Coulomb interactions to
\[
V(r)
=
\left\{
\begin{array}{ll}
\frac{\mu_0}{4 \pi} \frac{q_i q_j}{r_{ij}}
&
r_{ij} \neq 0
\\
v_0 \, q_i q_j
&
r_{ij} = 0
\,,
\end{array}
\right.
\]
where the value of the onsite Coulomb interaction $v_0$ can be chosen so that the dipolar
energy of two neighboring dipoles is exactly recovered in the dumbbell model. In fact we
can do better. There is also a small nearest neighbor exchange term in the Hamiltonian,
\begin{equation}
H
=
\frac{J}{3} \sum_{\langle ij \rangle} {\bf S}_i \cdot {\bf S}_j
\label{eq:exchH}
\end{equation}
and we can choose $v_0$ so that this is {\it also} included correctly~\cite{Castelnovo2008}:
\begin{equation}
v_0 \left(\frac{\mu}{a_d}\right)^2
=
\frac{J}{3}
+
\frac{4}{3} \left( 1 + \sqrt{\frac{2}{3}} \right) D
,
\label{eq:v0}
\end{equation}
where $D = \mu_0 \mu^2 / 4 \pi a^3$ is the dipolar coupling constant at the dipole-dipole nearest neighbor distance $a$.

\begin{figure}
\includegraphics[width=0.5\columnwidth]{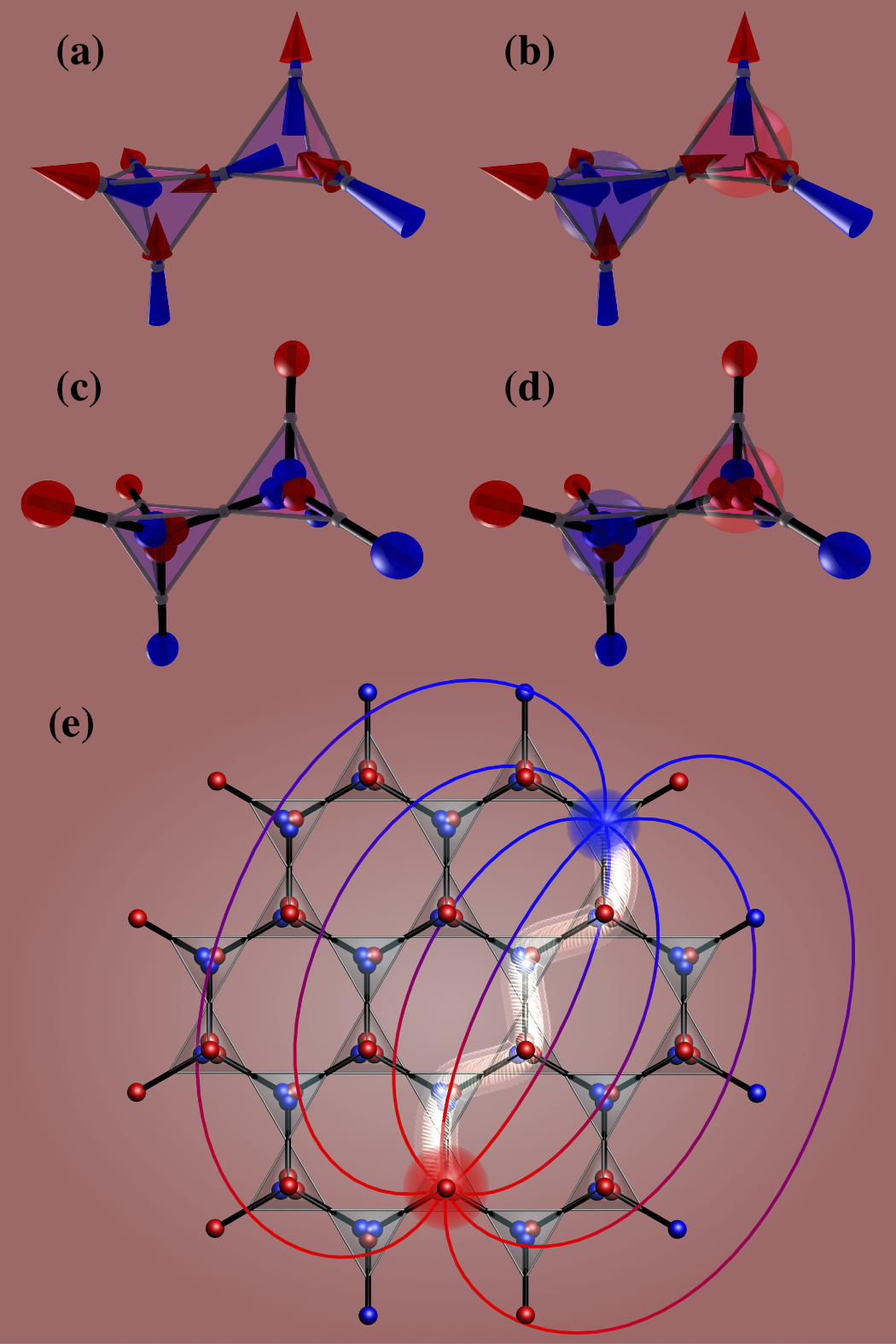}
\caption{
[From Ref.~\cite{Castelnovo2008}, Fig.~2]
Illustration of the spin to dumbbell mapping for ice rule satisfying
tetrahedra (a vs c) and for tetrahedra hosting positive and negative
monopoles (b vs d).
A ``Dirac string'' between two monopoles is shown in (e).
}
\label{fig:spinice_details}
\end{figure}

The net result of these replacements is that we can rewrite the energy function in terms of the {\it net} charges $Q_\alpha \equiv \sum_{i \in \alpha} q_i = 0, \pm 2 q_m, \pm 4 q_m$ at the centers $\alpha$ of the tetrahedra (which form a diamond lattice):
\begin{equation}
H =
\frac{\mu_0}{4 \pi} \sum_{\alpha < \beta} \frac{Q_\alpha Q_\beta}{r_{\alpha\beta}}
+
\frac{v_0}{2} \sum_\alpha Q_\alpha^2
.
\label{eq:dumbH}
\end{equation}
This equation encapsulates the dumbbell model. The unmodified Coulomb limit is recovered by
taking $v_0 \rightarrow \infty$; in that limit it is clear that the ground states of the
model consist of all configurations for which $Q_\alpha = 0$ for all $\alpha$. These are
exactly the ice rule satisfying configurations (2in-2out; Fig.~\ref{fig:spinice_details}a,c).
At finite $v_0$, the low-temperature fate of the system is determined by the competition
between the self-energy cost of a charge and the Madelung energy gain of an arrangement of
positive and negative charges on the diamond lattice.
One can check that the ground state set of Eq.~(\ref{eq:dumbH}) remains unchanged when
$v_0$ is sufficiently large, which is the case for physical values in {\dys} and {\holm}.
We have thus shown that the dumbbell
model---which captures most of the dipole interaction accurately including its long ranged
part---has {\it exactly} the same ground states as the nearest neighbor model of spin ice.

We would like to ask  the reader take a moment to admire what has been accomplished. For a model
of hard (fixed length) spins with genuinely long ranged and frustrated interactions, we have found
$O(e^N)$ exact ground states. Normally, even finding {\it one} such ground state would be a
tall order.
Here, the lattice anisotropy and our choice of interactions have conspired to make the task trivial.
What is remarkable is that there are actual compounds, the spin ice materials, that realize this
model to an excellent approximation.
Conversely, the dumbbell model, whose energetics differ from that of the purely dipolar model by
manifestly small terms, explains why spin ice exhibits the Pauling entropy at low
temperatures.
In principle, the deviations from the perfect degeneracy captured in the dumbbell model do give rise to ordering as $T \rightarrow 0$~\cite{Siddarthan1999,Bramwell2001,Isakov2005}.
However, these appear to be experimentally inaccessible---the system freezes before any ordering
is detected---and they are thus irrelevant in the discussion of the actual physics of these
compounds.
%
%

\section{\label{sec: monopoles}
Monopoles
        }
Now let us turn to the excitations. As far as their energetics is concerned, we can again work with
the dumbbell model. The simplest move out of the ground state manifold consists of flipping a
single spin which breaks the ice rule for two neighboring tetrahedra as illustrated in Fig.~\ref{fig:spinice_details}b,d.
In the dumbbell model, the corresponding move creates two equal and opposite charges $\pm 2 q_m$
on nearest neighbor diamond sites. Now these charges do not {\it have} to sit next to each other---they
can be moved apart by flipping a sequence of spins/dumbbells as illustrated in Fig.~\ref{fig:spinice_details}e. In other
words, the initial spin flip can fractionalize into two defect tetrahedra which can move independently.
A key question in any such construction is the energetic feasibility of the fractionalization, i.e.
are the fractionalized objects (``quarks'') deconfined?~\cite{foot4}
In our case the answer is immediate thanks
to the dumbbell model wherein the energy of two defects located a distance $r$ apart is simply
\[
E(r)
=
2 \frac{2 v_0 \mu^2}{a_d^2} + \frac{\mu_0}{4 \pi} \frac{(2 \mu / a_d)(-2 \mu / a_d)}{r}
\]
which is the sum of two defect creation energies $\Delta=2 v_0 \mu^2 / a_d^2$ and a (magnetic) Coulombic interaction between the defects. Given the nature of the magnetic interaction between the defects
it is appropriate to call them monopoles, but more on that anon.
%
%

\subsection{
Two component plasma and Debye-H\"{u}ckel theory
           }
Our considerations above are summarized by the formulation that we have mapped the energetics of the
low energy configurations of spin ice onto that of a set of monopoles and anti-monopoles
with a finite creation energy per particle and a Coulomb interaction between them---a
system that has been known and studied for many years and goes under the name, among others, of the two component
plasma~\cite{Levin2002}. While this mapping requires some further discussion and qualification to which we will return below, let us first show that it provides an simple yet accurate account of the low temperature thermodynamics of spin ice.

\begin{figure}
\includegraphics[width=0.5\columnwidth]{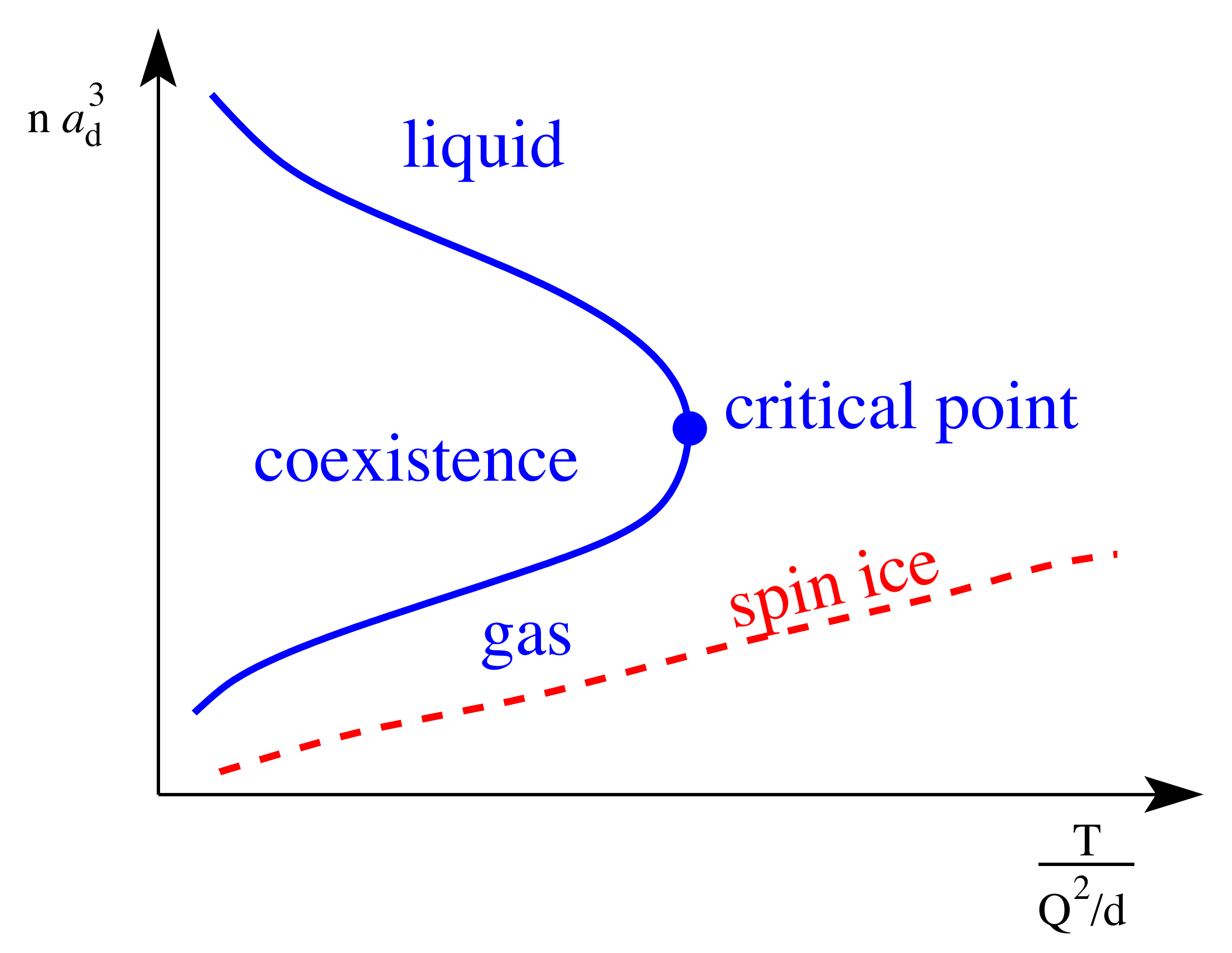}
\caption{
Sketch of the generic phase diagram of a two component plasma.
In zero field, spin ice traces a trajectory as a function of temperature that lies entirely in the gaseous phase.
}
\label{fig:liq-gas}
\end{figure}

The natural parameters in the plasma are the average thermal energy of the particles $k_B T$,
the magnitude $Q = 2 q_m = 2\mu/a_d$ of their charges, and their average separation defined by their density $d \sim n^{-1/3}$. In addition the plasma needs a short distance cutoff to regulate the Coulomb attraction, absent which the system suffers collapse. In our problem that is naturally provided by the lattice constant $a_d$.
From these quantities we generate two dimensionless ratios that control
the physics of the system. One can choose these to be the plasma parameter or the interaction strength in units of the temperature $\Gamma = {Q^2/d \over k_B T}$ and dimensionless density in units of the short distance cutoff $\tilde{n}= n a_d^3$. In terms of these, the plasma exhibits a phase diagram with gas, liquid and coexistence
regions topped by a critical point as sketched in Fig.~\ref{fig:liq-gas}.

The monopole density in spin ice varies with $T$; at asymptotically low temperatures, it vanishes as
$\tilde{n} \sim e^{-\Delta/T}$. At the same time the dimensionless interaction strength also decreases with temperature. It follows that spin ice traces a path on the phase diagram in Fig.~\ref{fig:liq-gas} that lies entirely
within the gaseous region, as sketched there. Note that this statement is equivalent to our earlier
statement that spin ice exhibits no phase transition as it is cooled from the paramagnetic phase.

A very useful understanding of this gaseous region can be gained by resorting to the standard
approximate treatment of dilute plasmas, namely Debye-H\"{u}ckel (DH) theory~\cite{Debye1923,Levin2002}.
Given the correct identification of the monopole charge, chemical potential, and lattice entropy appropriate for spin ice, it is straightforward to use the DH equations in order to compute the heat capacity.
This calculation can be compared to the experimental heat capacity data by Klemke and collaborators
(see Fig.1b in Ref.~\cite{Morris2009}), and it is apparent that DH theory provides a good
understanding of the low temperature behavior in the regime where monopoles are sparse.
It is also clear that this plasma computation does
much better than traditional spin system methods such as cluster expansions of the free energy
or Cayley tree approximations and thus provides evidence for the correctness of the plasma formulation.

A further tunable knob is needed in order to access the rest of the plasma phase diagram. Here the
specific structure of the local easy axis anisotropies on the pyrochlore lattice comes to the rescue: while spin
ice is an antiferromagnet in pseudospin language, a magnetic field couples to the `real' magnetic moments and can
thus lift degeneracies as the moments do not locally sum up to zero even in the ground states~\cite{Moessner1998}.

\begin{figure}
\includegraphics[width=0.5\columnwidth]{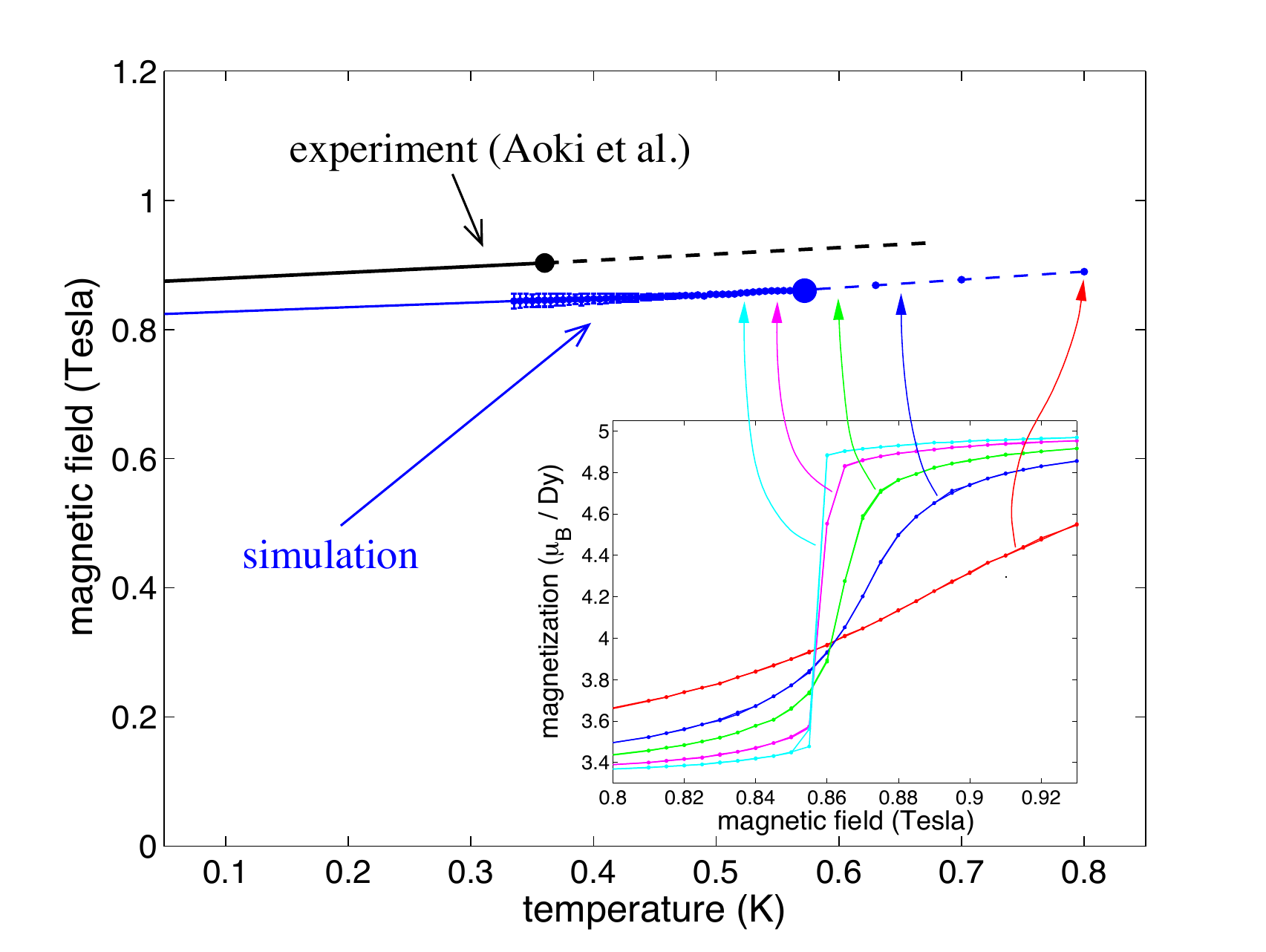}
\caption{
[From Ref.~\cite{Castelnovo2008}, Fig.~4]
Phase diagram of spin ice as a function of temperature and magnetic field applied in the [111] direction. The inset illustrates the behavior of the magnetization above and below the critical
temperature (from Monte Carlo simulations). 
The experimental results are from Ref.~\onlinecite{Aoki2004}. 
}
\label{fig:phdiag}
\end{figure}

One thus observes (see Fig.~\ref{fig:spinice_details}a-b) that applying an external magnetic field along one of the [111] axis
eventually favors a (unique) configuration with only 1in-3out and 1out-3in tetrahedra.
As such, it acts as
a chemical potential for the monopoles, which can now be tuned independently of temperature~\cite{Castelnovo2008}. This predicts that the phase diagram in the $(H,T)$ plane should
exhibit a first order gas-liquid transition line terminated by a critical end point. In fact,
exactly this phase diagram was observed in spin ice prior to the formulation of the
theory~\cite{Sakakibara2003} and already then flagged as being unusual.
The first order line exhibits a jump in magnetization, which is the same in
this context as a jump in the monopole density (see Fig.~\ref{fig:phdiag}),
and the jump goes away at a critical point.
%
%

\subsection{
Monopoles with strings attached
           }
Everything we have said thus far about the monopoles has only turned on their Coulombic interaction and
that is an unproblematic consequence of their energetics, as most transparently encoded in the dumbbell
model. We turn now to a more careful consideration of the objects themselves.

We begin with a generality---while a condensed matter system can produce objects that act as fractionally
charged sources for the electric field $\mathbf{E}$, it cannot do the same for the magnetic field $\mathbf{B}$.
This distinction holds for long wavelength fractional charges, which arise from averaging over microscopic sources. While $\mathbf{\nabla} \cdot \mathbf{E}= \rho$ and averaging can give rise to a fractional source
on the RHS, $\mathbf{\nabla} \cdot \mathbf{B}= 0$ and no amount of averaging can producing anything except zero
on the RHS. Thus the apparent consequence of the dumbbell model, that the monopoles are sources of
$\mathbf{B}$, cannot be correct.

It is easy to see where the problem lies. In introducing dumbbells we introduced degrees of freedom which {\it are} microscopic sources of $\mathbf{B}$. While this was unproblematic in the ground
states, it becomes problematic in the excited states. Going back to the original dipoles remedies
that problem. In the dipole language we create two separated monopoles starting with a given
ground state by identifying a chain of dipoles arranged head to tail and flipping them. This
is equivalent to starting with the initial configuration and adding two copies of the reversed
dipole chain. {\it Relative} to the starting ground state we have introduced the magnetic field
of a dipole chain with a dipole moment density $-2 \mu/a_d$ per unit length. The magnetic field
$\mathbf{B}$, due to this inserted object, is that of a monopole-antimonopole pair at its endpoints
{\it and} a ``Dirac string'' of flux connecting them along the dipole string (see Fig.~\ref{fig:spinice_details}e). The
existence of the string now accounts for the solenoidal constraint.

Three comments on the string are in order. First, its existence evades the classical Dirac argument
for monopole quantization. Indeed, in spin ice the monopole charge is roughly 8000 times smaller
than the Dirac value~\cite{Dirac1931}---in part testimony to the smallness of the fundamental dipole, which itself
is forced to arise from currents, and in part a consequence of fractionalization on the scale
of the lattice constant. Second, the deconfinement of the monopoles finds an alternative to the
classical Dirac construction---in which the Dirac string is made unobservable---by making the
string tensionless and thus infinitely extensible. Third, there is really no string, in that
it can only be defined with respect to a particular starting ground state and it changes if one
changes the reference ground state. If simply handed a spin ice configuration with monopoles, one
cannot uniquely identify strings in it. This ambiguity also feeds into the feature
that the actual spatial distribution of $\mathbf{B}$ for a monopole in a given ground state, which
arises from all moments in the system, is not particularly monopolar except under special circumstances
such as the ``Stanford monopole experiment" mentioned below.
 Nevertheless, we shall see shortly that one can experimentally observe strings if we force the system
 into a given ground state initially.

Finally, we note that the monopoles can be identified as sources of the magnetic field$\mathbf{H} = \mathbf{B} - \mathbf{M}$ after appropriate coarse graining. Specifically, consider the smeared magnetic charge
\[
\rho_m(\mathbf{R})
=
\int d^3\mathbf{r} \:
  \exp\left(
    -\frac{\vert \mathbf{r} - \mathbf{R} \vert^2}{\xi^2}
  \right)
  \mathbf{\nabla}\cdot\mathbf{H}
\]
In the limit $\xi \gg a$ this returns $\pm 2q_m$ for isolated magnetic monopoles/antimonopoles.
%
%

\subsection{
Imaging the Dirac strings
           }
As explained above, creating a separated monopole-antimonopole pair
requires flipping a chain of head-to-tail aligned spins. In general,
this Dirac string is statistically indistinguishable from the ice rule obeying background and no (local) spin
correlation function can detect it.

This statement holds for typical ice rule satisfying configurations---the vast majority,
in fact---but not all of them. Indeed, Morris and collaborators~\cite{Morris2009} took
advantage of a notable exception and elegantly succeeded in measuring the Dirac strings, thus
providing indirect confirmation of the monopole excitations in spin ice materials.

The key step at the root of this experiment is to use a [001] field to bring the system to
magnetic saturation, which can be achieved with fields much weaker than the anisotropy.
This selects a {\em unique} configuration in which all spins have a positive projection onto
the field direction, and thus minimise the Zeeman energy.
From Fig.~\ref{fig:pyro} one can immediately see that this state also satisfied the ice rules.

Before we turn to the role of monopoles, we briefly digress to note that the transition between
this state and the spin ice states with unsaturated magnetization reflects the topological
character of the states satisfying the ice rules: the magnetization of each [001] plane is the
same throughout the lattice. Therefore, as the field is reduced, the magnetization can only
be reduced  in the absence of monopoles by a `non-local' move, namely by flipping a string
of spins spanning the entire system: in the saturated state, no local fluctuations are possible!
The transition out of this state therefore has the unusual character of appearing fluctuation-free
on the high-field side, while looking like a perfectly standard second-order transition on the
low-field side, because an entropic repulsion between the strings leads to a gradual lowering of the
magnetization~\cite{Jaubert2008}. This transition bears Kasteleyn's name, who first discussed
this behaviour in the context of planar dimer models.

In experiment, such non-local spin flips are not possible. Instead, reducing the field at low temperature
generates a dilute set of monopole-antimonopole pairs.
These pairs are connected by finite-length Dirac strings, defined relative to the
saturated starting state. This string {\it must} stretch monotonically in the [001] direction although it
is free to meander transversely on the lattice---indeed, it meanders randomly in the
transverse plane.

\begin{figure}
\includegraphics[width=0.5\columnwidth]{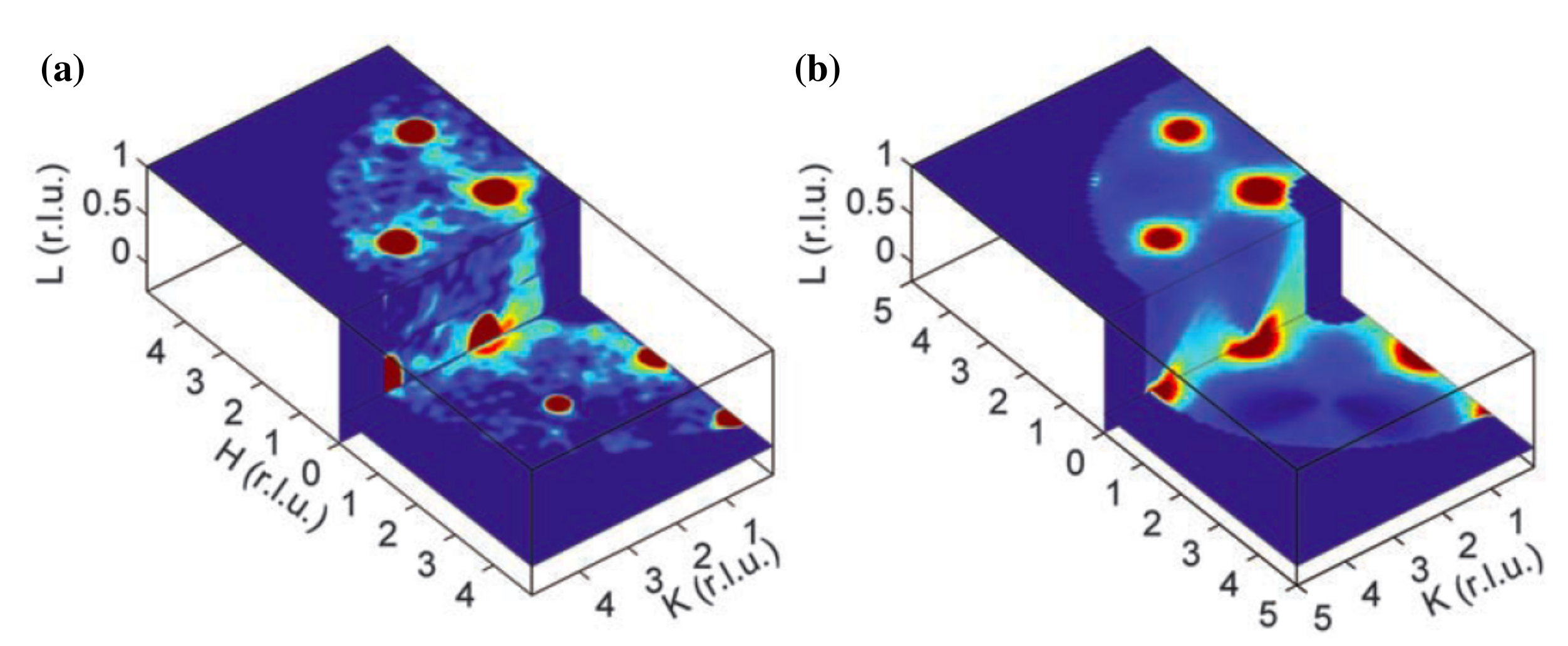}
\caption{
[From Ref.~\cite{Morris2009}, Fig.~3]
Neutron diffraction data showing a cone of scattering appearing at the (020) Bragg peak (a),
compared with the result calculated using two-dimensional random walk considerations (b).
}
\label{fig:morris}
\end{figure}

The diluteness of the Dirac strings allows them to be observed. More precisely, they provide a set
of dilute and thus independent degrees of freedom which can be used to generate a
compact description of the scattering pattern. The remaining task is to compute the
scattering due to a single string and here its orientation comes to our aid. Thanks
to the latter, we are able to view the strings as two-dimensional random walks in the $xy$
plane, drawn as a function of ``time'' in the $z$ direction. The correlation between two spins
separated by a vector $(x,y,z)$ is directly
related to the probability that a two-dimensional random walk starting at the origin
passes by position $(x,y)$ after time $z$.
On the pyrochlore lattice, with spins pointing along the local [111] easy axis, these correlations
translate into characteristic cone-shaped intensity patterns in the structure factor, which are
clearly observed in diffuse neutron scattering measurements~\cite{Morris2009}, as
illustrated in Fig.~\ref{fig:morris}.
Tilting the field away from the [001] direction gives a bias to the random
walk, as the string prefers to visit those sites which are cheaper to flip against the
applied field; this was again observed experimentally (not shown).
%
%

\section{
Emergent Gauge Field
        }
In the discussion thus far we have talked about a low temperature description entirely in terms of the
energetics of the monopoles. And yet this cannot be the whole story as we have ignored one of the
defining features of spin ice, its macroscopic zero temperature entropy. This entropy turns out to
have two consequences. First, the zero temperature limit of the spin correlations is nontrivial---it involves
an average over all the ground states. Second, the spin entropy is modified by the presence of monopoles
and it is sensitive to their location. In turn this results in an entropic force between the monopoles.
Interestingly, both effects are captured by an emergent gauge field governed by a Maxwell action~\cite{Huse2003,Moessner2003b,Isakov2004,Hermele2004,Henley2005,Henley2010}.

Consider the spin ice ground states, with the ice rule forcing two spins to point in and two to point out
on each tetrahedron. If we change our perspective, we can think of these as the configurations of a lattice
vector field $\mathbf{S}$ that lives on the bonds of the dual diamond lattice, takes values oriented along and opposite to
each bond {\it and} satisfies the lattice discretization of the conservation law $\mathbf{\nabla} \cdot \mathbf{S(x)}=0$.
If we now switch to a somewhat coarse grained description, we go from weighting all states equally to
generating a probability distribution,
\begin{equation}
P[\mathbf{S}(\mathbf{x})]
\propto
\delta(\mathbf{\nabla} \cdot \mathbf{S}) \,
e^{-\frac{K}{2} \int d^3\mathbf{x} \: \left\vert \mathbf{S}(\mathbf{x}) \right\vert^2}
.
\label{eq:entropicS}
\end{equation}
The particular form here follows from a) the requirements of analyticity and minimum
scaling dimension for the output of a short distance coarse graining (a standard renormalization
group argument), b) the observation that microscopic configurations with lots of short loops of
the lattice vector field dominate the ground state sum and coarse grain to configurations with
small values of $\mathbf{S}(\mathbf{x})$, and c) the requirement
that the coarse graining preserve the solenoidal constraint. [Note that the coarse
grained spin field is proportional to the coarse grained magnetization $\bf M(x)$.]

\begin{figure}
\includegraphics[width=0.5\columnwidth]{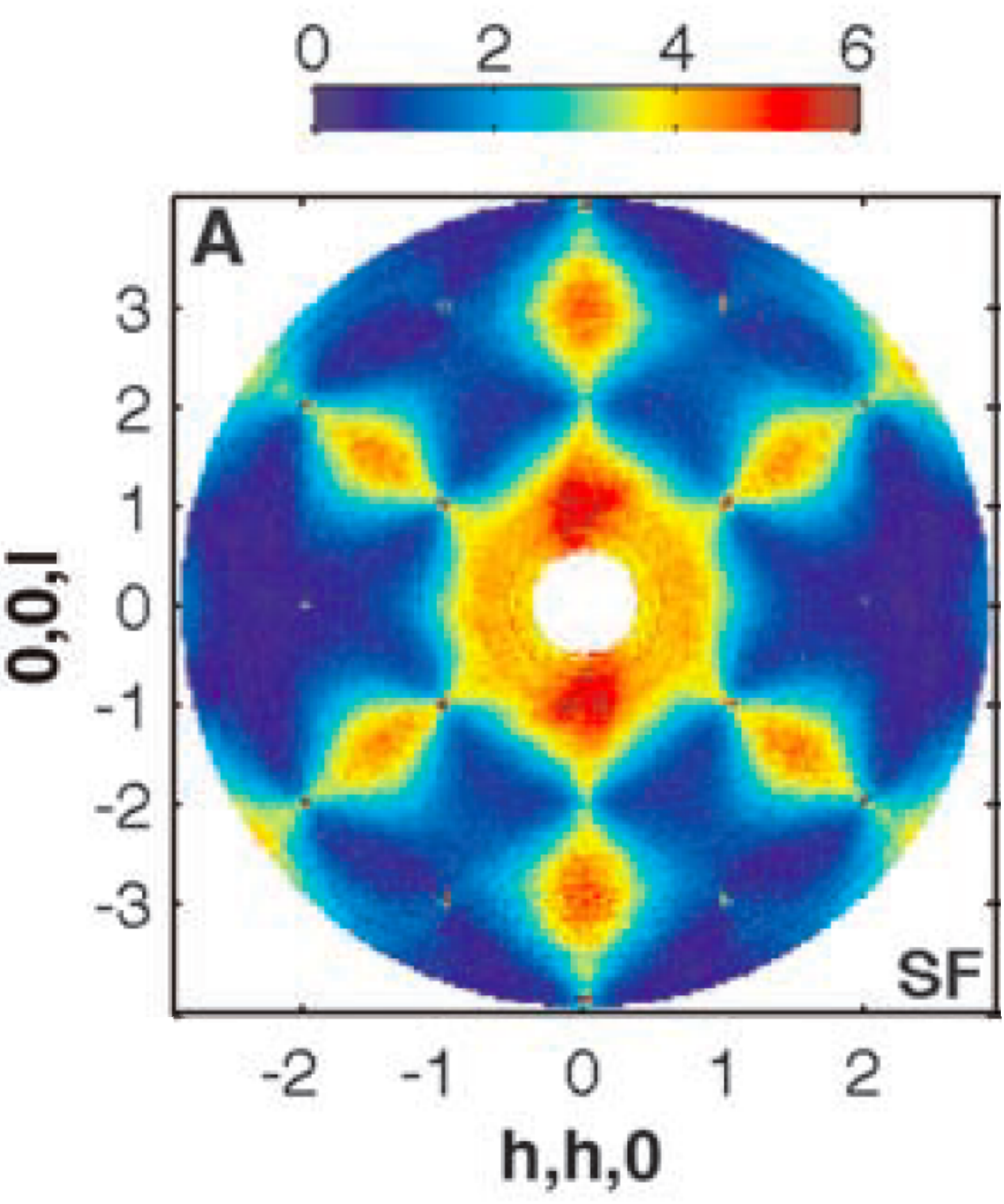}
\caption{
[From Ref.~\cite{Fennell2009}, Fig.~2A]
Diffuse scattering maps from spin ice {\holm}.
The dipolar nature of the spin ice correlations, Eq.~(\ref{eq:dipcorr}), lead to a characteristic
$\cos^2(\theta)$ angular dependence of the intensity of the structure factor in reciprocal space:
as one moves around a Brillouin zone centre, say (0,0,2), the intensity goes through a maxima
(red) and minima (blue) twice as we complete a full 360$^{\rm o}$ contour.
}
\label{fig:pinchpoints}
\end{figure}

At this point it is trivial to solve the
constraint via the introduction of a gauge field
$\mathbf{S}(\mathbf{x}) = \mathbf{\nabla} \times \mathbf{A}(\mathbf{x})$ and obtain
the probability distribution of a Euclidean Maxwell theory,
\[
P[\mathbf{A}(\mathbf{x})]
\propto
e^{-\frac{K}{2} \int d^3\mathbf{x} \:
  \left\vert \mathbf{\nabla} \times \mathbf{A}(\mathbf{x}) \right\vert^2} \ ,
\]
thus uncovering the promised gauge field.
It will now not surprise the reader that this analysis implies that the long distance spin
correlations are dipolar~\cite{Isakov2004},
\begin{equation}
\langle S_i({\bf x}) S_j({\bf 0}) \rangle
\propto
\frac{3 x_i x_j - r^2 \delta_{ij}}{r^5}
. 
\label{eq:dipcorr}
\end{equation}
In turn, the correlations imply the existence of ``bow-ties'' or pinch-points in the neutron scattering
pattern in momentum space. Gratifyingly, such singularities have indeed been
observed~\cite{Fennell2009,Kadowaki2009} in low temperature neutron
scattering on spin ice (Fig.~\ref{fig:pinchpoints}).
Readers should note that these dipolar correlations are {\it not} a
trivial consequence of the dipolar interaction---after all they exist already for the short ranged
model of spin ice. Of course there {\it is} a deeper connection---it is the happy coincidence of the
entropic and energetic dipolar correlations that makes the physics of spin ice stable to the
inclusion of the full range of the dipolar interaction.

A second consequence of the emergent gauge field is that the monopoles, as sites that violate
the solenoidal constraint, are also sources of its flux and thus experience the standard Coulomb
interaction with a strength set by $KT$, the latter factor accounting for the translation between
the entropy encapsulated in Eq.~(\ref{eq:entropicS}) and the free energy. Thus there is this
elegant feature that the full interaction between monopoles is still Coulombic---now with a
$T$ dependent strength~\cite{DHpaper}. [Historically, the entropic interaction was
understood~\cite{Huse2003,Moessner2003b,Isakov2004,Hermele2004,Henley2005} prior to the identification of the energetic interaction.] For
practical purposes though, the energetic part dominates strongly at the relevant temperatures
in the current compounds although that could change in spin ice realizations with smaller moments
and thus (relatively) stronger nearest-neighbour exchange.

The thermally excited monopoles do however cut off the dipolar correlations and thus define
a length scale---their separation---that functions as a divergent correlation length as $T
\rightarrow 0$. Thus we find that as $T \rightarrow 0$ spin ice approaches a critical point,
characterized not by the development of quasi-long ranged order in an order parameter but
instead by the deconfinement of a gauge field. While this is thermal physics, it is in
remarkably close analogy with Polyakov's explication of the physics of compact $U(1)$
gauge theory near the Maxwell limit in 2+1 dimensions~\cite{Polyakov1977}.

Finally, we should note two related instructive aspects of spin ice physics. First, the
feature that monopoles are sources of both the standard magnetic field and of the emergent
gauge flux is common in topologically ordered systems, although it was not recognized until
it cropped up in spin ice~\cite{Moessner2010}: it has the consequence that one must distinguish the charge
under the emergent gauge field, which is quantized, from the charge under the magnetostatic gauge
field, which is not. This distinction generalizes to other topologically ordered systems, in particular to
models of electric charges on the pyrochlore lattice,~\cite{Fulde2002}, or indeed to water ice itself.

Second, the application of a [111] magnetic field leads to a magnetization plateau in which
there is still a residual macroscopic entropy. The entropy of this ``kagome ice'' state~\cite{Matsuhira2002,Udagawa2002,Higashinaka2003,Moessner2003} comes from purely planar
rearrangements of the spins and it is captured by a two dimensional critical gauge field, more commonly
known as a height field. Thus we have the interesting phenomenon of a dimensional reduction of
the emergent gauge structure by the application of an external field.

The fluctuations of this
height field can be quenched by tilting the field towards the [$\overline{\rm 1}\overline{\rm 1}{\rm 2}$]
direction and the
resulting two dimensional Kasteleyn transition~\cite{Moessner2003} has also been observed~\cite{Fennell2007}.
%
%

\section{
Dynamics
        }
Thus far we have discussed the low temperature statics and thermodynamics of spin ice. Let
us now turn to what is known about its dynamics in and out of equilibrium. We warn the
reader at the outset though that this is a subject that is not nearly as well settled as the
thermodynamics and statics.

Let us begin with two overarching observations. The first is empirical: relaxation times in spin ice
grow extremely rapidly at low temperatures and below about 600mK spin ice is generally not in equilibrium
on laboratory time scales. As evidence see the contrasting measurements of field-cooled and
zero-field-cooled magnetization and the measurements of the uniform magnetic
susceptibility~\cite{Matsuhira2000,Snyder2004}. The second observation is theoretical: the
low temperature dynamics of
spin ice is sensitive, as befits a classical system, to various details that do not affect the
equilibrium behavior. These include the relative magnitudes of processes that move the system in
the ground state manifold and processes that involve the creation, motion and recombination of
monopoles as well as the microscopic time scales for attempting such transitions. Our current
best understanding is
that the low temperature dynamics of spin ice below about 10K is well captured by a stochastic,
``Monte Carlo'', process in which single spins attempt to flip about every millisecond at a
relatively $T$ independent attempt frequency. This has the consequence that while
existing monopole motion is fast, their pairwise creation
is a slow process and there is no significant direct motion in the ground state manifold,
via ``ring exchange'' terms. It is nonetheless important to remark that such ring exchange terms
are present in principle, and they could be boosted via local chemistry, e.g.\  in compounds
with a smaller spins or, more generally, a different crystal field level scheme.
%
%

\subsection{
Equilibrium dynamics
           }
\begin{figure}
\includegraphics[width=0.5\columnwidth]{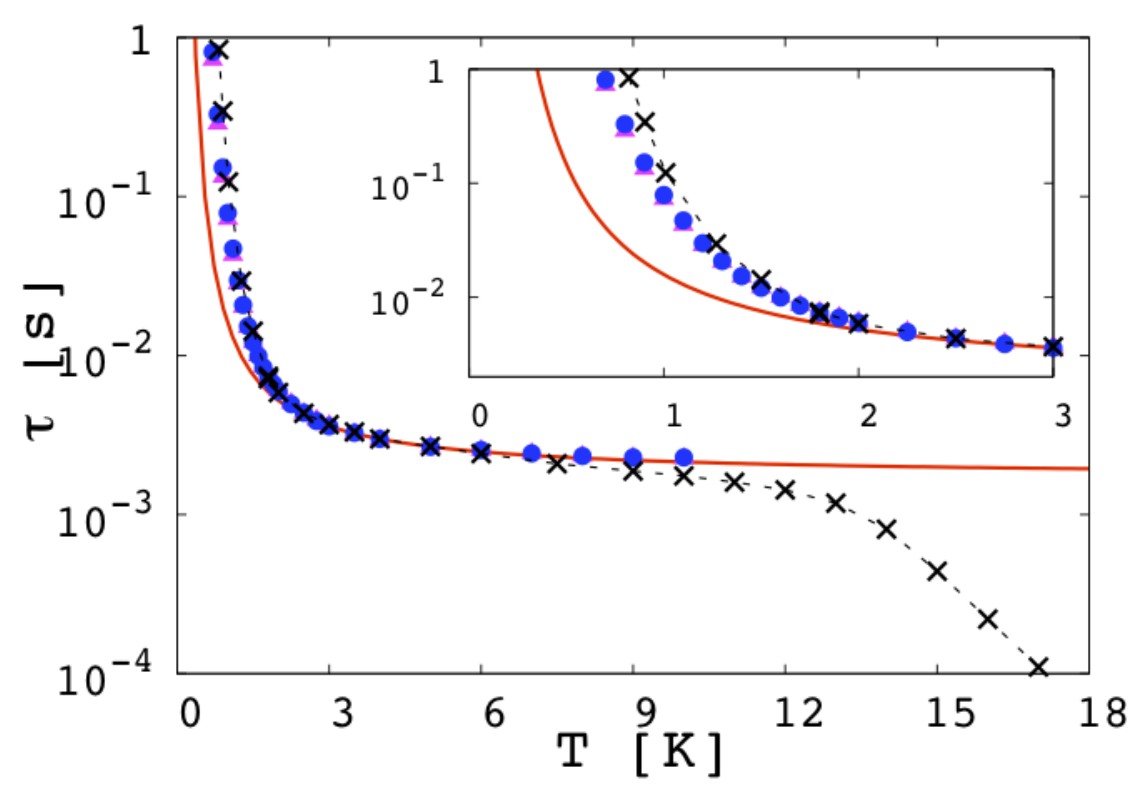}
\caption{
[From Ref.~\cite{Jaubert2009}, Fig.~2]
Experimental magnetic relaxation time scale $\tau$ as a function of temperature
from susceptibility data in Ref.~\cite{Snyder2004} (black crosses and dashed line).
The rapid increase in $\tau$ at low temperatures is due to the paucity of
defects responsible for the magnetic rearrangement of a spin ice configuration
(namely, the monopoles).
This increase cannot be described by the Arrhenius law that obtains in the non-interacting
approximation (red line).
On the contrary, the spin autocorrelation time scales computed using Monte Carlo simulations
in Ref.~\cite{Jaubert2009} are in good agreement with the experimental data.
To leading order, the improvement is due to a finite screening length that reduces the
isolated monopole cost at intermediate temperatures.
As discussed in Ref.~\cite{DHpaper}, this effect of the interactions on the time scales can
be captured using Debye-H\"{u}ckel theory.
}
\label{fig:snyder}
\end{figure}

Ideally, we would like to have an understanding of the frequency and wave-vector dependent spin correlations at low temperatures. Experimentally, what is available at present is the  $q=0$
small $\omega$ response measured in AC susceptibility measurements~\cite{Matsuhira2001,Snyder2004}.
These authors extracted a timescale that exhibits a rapid
increase as the temperature is decreased, quickly becoming larger than the accessible time window
(Fig.~\ref{fig:snyder}).
The increase was so rapid that they argued it could be faster than exponential. We will now
describe the current state of the theory.

The first treatment of the dynamics in spin ice in terms of the gauge field/monopole description
was given in Ref.~\cite{Moessner2003} for the kagome ice plateau in a [111] field discussed
above, which relied on collective dynamics for the gauge field generated by ring exchange terms, i.e. cooperative moves of the
spin that respect the ground state conditions. As we have noted above, it appears that the
ring exchange terms are too small at accessible temperatures. [It is perhaps worth
emphasizing that this is a quantitative issue for the collective dynamics will
dominate as $T \rightarrow 0$ for an ideal spin ice system.]

The next step was taken by Ryzhkin in a paper~\cite{Ryzhkin2005} which did not get the
attention it deserved at the time due to an abstract that announced a noninteracting treatment of the monopoles. Although he incorrectly identified the decay of the entropic forces as being faster than a Coulomb interaction, he did come close to identifying the energetic Coulomb force. Importantly, he did recognize the role of the monopoles as sources and sinks for the magnetisation and their significance for magnetic relaxation in the low-temperature, hydrodynamic, regime. 
In this regime the relaxation of the magnetization by monopole motion
is summarized by the equation~\cite{Ryzhkin2005}
\[
\frac{\partial \bf M}{\partial t} = {\bf j_m}
\]
and the constitutive relation
\[
{\bf j}= - 2 \mu n \Phi {\bf M}
\]
where $\mu,n$ are the monopole mobility and density and $\Phi= (8/\sqrt{3}) a k_B T$. The
$T$ dependence of $\Phi$ indicates the entropic origin of the monopole drift in response
to the local magnetization~\cite{Ryzhkin1997}. Together these two equations imply a relaxation time
\[
\tau \sim \frac{1}{2 \mu n \Phi} \ ,
\]
which diverges exponentially $\sim e^{\Delta/T}$ at low temperatures due to the vanishing monopole density, as well as a Debye relaxation form for the susceptibility. It seems to us that this analysis is likely asymptotically correct as $T \rightarrow 0$ at $q=0$, although it needs to be supplemented
by the explicit inclusion of Coulomb effects in order to treat $q \ne 0$.

A comparison between this non-interacting exponential increase and the experimental results in
Ref.~\cite{Snyder2004} leads however to an unsatisfactory agreement at accessible temperatures.
A much better fit is obtained if the monopole density is computed using DH theory~\cite{DHpaper}
indicating that Coulomb effects are significant for the measurements even at $q=0$.
An even better fit to the data has been obtained by Jaubert and Holdsworth~\cite{Jaubert2009}
who have computed a relaxation time from the spin autocorrelation function
by explicit Monte Carlo (MC) simulations of dipolar spin ice.

As noted above, the dynamics of spin ice is believed to be a good approximation to MC dynamics.
This is not a trivial problem as one still needs to deal with the serious limitations on the system sizes
that can be simulated, imposed by the computational cost of treating the long range dipolar
interactions. These are particularly crippling at low temperatures, when the density of
monopoles becomes exponentially small.
In Ref.~\cite{Jaubert2009}, the problem was partially solved with a clever approximation:
an algorithm that retains the full spin configuration but computes its energy using the
effective monopole description, in terms of chemical potential and long range Coulomb interactions.
Contrary to the conventional dipolar approach, this new algorithm becomes more and more
efficient as the temperature is decreased and the number of long-range interacting terms is
exponentially suppressed---although it still has to deal with the need to simulate an
ever larger system.

We note that while the results of Ref.~\cite{Jaubert2009} fit the data of Ref.~\cite{Snyder2004}
quite well (see Fig.~\ref{fig:snyder}), there are two unresolved issues in this agreement.
First, it is not yet possible to access via MC the putative transition to the naive exponential growth
that would serve as a test of its asymptotic agreement with the hydrodynamics as $T\rightarrow 0$.
Second, the local spin autocorrelation function is an integral over all $q$ and that is not what
is measured in Ref.~\cite{Snyder2004}.

This last point brings us to the observation that the dynamics at low $T$ are also dynamics
near a critical point with a divergent correlation length and thus should also fit into
the framework of dynamical critical phenomena~\cite{Hohenberg1977}.
The related problem of
multicomponent antiferromagnetic spins on the pyrochlore lattice has been shown to
exhibit dynamic scaling~\cite{gregorthesis,Conlon2009} for long wavelengths and
low frequencies. The spin ice problem still awaits a definitive analysis along these
lines.
%
%

\subsection{
Out of equilibrium behavior
           }
The most obvious reason to be interested in the behavior of spin ice out of equilibrium is that
this is what is observed experimentally at low temperatures where time scales of order of a
few days and longer are easily arranged. Second, there is the more fundamental question of
how an ensemble of pointlike defects embedded in a network of strings behaves. Finally,
there is the enticing prospect of constructing by
means of magnetic charges analogs of familiar circuit phenomena with electric
charges.

Ref.~\cite{Castelnovo2010} considers the fate of the system following a thermal quench from high
temperatures where monopoles are abundant down to low temperatures where they are sparse.
As monopoles can only pair annihilate, one obtains a diffusion-annihilation problem, in which
the monopole density decays slowly as each monopole first needs to move to find a partner
with which to annihilate. Whereas the magnetic Coulomb force helps in locating such a partner,
it turns out that at low temperature it in fact slows down the annihilation process as a result of a
simple trapping mechanism: flipping the spin shared by two tetrahedra hosting a pair of
oppositely charged monopoles annihilates this pair only if this spin flip repairs the ice rules; if instead
it leads to the creation of charge-2$Q$ monopoles, the monopoles are stuck as there is a Coulomb
energy barrier for them to annihilate by flipping a longer string of spins~\cite{Castelnovo2010}. The
counterintuitive fact that this dynamical arrest is most visible at low temperature when the equilbrium
monopole density is lowest reflects an unusual interplay of lattice-scale and long-wavelength degrees
of freedom.

The magnetic two-component Coulomb plasma analogy outlined above has been pushed
beyond equilibrium phenomena in an imaginative
stream of work by  Bramwell and co-workers~\cite{Bramwell2009,Giblin2010} aiming to transpose
concepts from the field of electrolytes to the study of spin ice; in this context the labels `magnetolyte'~\cite{CastelnovoCPC}
and `magnetricity'~\cite{Bramwell2009} have been applied to spin ice and  the collective behavior of
the monopoles it hosts.
They examine the response of spin ice to a small magnetic field, specifically analyzing the experiments
in terms of Onsager's modeling~\cite{Onsager1934} developed in the framework of the Wien effect,
which refers to the change in analogous electrical response properties of an electrolyte. This
work turns on managing the complexity of length scales in the plasma by truncating it to a binary
classification between bound and free monopoles and writing rate equations for the latter. While
the soundness of various approximations employed in this work remains to be established, its
success in modeling data suggests that it is indeed possible to give a coherent account of
spin ice even out of equilibrium in terms of the monopoles and their interactions.

For a complete understanding of the non-equilibrium behavior of spin ice, it has become apparent
that a model in terms of spins only will not be sufficient. Demagnetization~\cite{Orendac2007} and
field sweep experiments~\cite{Slobinsky2010,Petrenko2011}
have unearthed a rich phenomenology with several different timescales.
For instance, at low temperatures, there are three distinct magnetization curves as a function of
field sweep rate. For (in practice, unattainably) low sweep rates, one on general grounds expects
to find the equilibrium magnetization curve. At intermediate sweep rates, the magnetization does not start growing until
a `critical' field is reached, when the Zeeman energy gain surmounts the smallest energy barrier for a spin
flip, whereafter a smooth growth in magnetization up to saturation occurs. For fast sweeps, plateaux in the
magnetization connected  by  discontinuous jumps occur; a jump goes along with heating of the sample,
indicative of an inability of the phonons to carry away the heat generated in a magnetic avalanche. Why
these avalanches appear around the same temperature where Monte Carlo simulations {\em not incorporating
phonons} find the rapidly growing timescales~\cite{Jaubert2009}
remains for the moment a mysterious coincidence.

Finally it is also worth noting that monopoles have been studied in artificial
spin ice in two dimensional lithographically patterned micromagnetic arrays since the
early work in Refs.~\cite{ASIref1,ASIref2}.
Here the dynamics is entirely non-equilibrium and poses, for example, the challenge of
understanding the role of monopoles in avalanche dynamics of these systems; for a popular account, see~\cite{Heyderman2011}.
%
%

\section{
Looking forward
        }
We hope we have succeeded in conveying to the reader the significance of the spin
ice compounds in the contexts of fractionization and topological order that we
highlighted at the outset---not least because they still are a rare, and exceptionally
transparent, realization of these ideas outside the setting of two dimensions and
high magnetic fields.

Our understanding of the current spin ice compounds is least complete in relation to their
dynamics and we sketched some open questions above. There is also the challenge of single
monopole detection, perhaps along the lines of the celebrated ``Stanford'' monopole search
as proposed in~\cite{Castelnovo2008}. Here we would detect the actual magnetic flux through a loop external to a long sample---a special setup that evades our earlier caveats about the
difficulty of detecting the magnetic flux associated with monopoles in general.

Clearly, it would be interesting to move beyond the
focus on {\dys} and {\holm}.
For example, a lowering of the gap to higher crystal field levels should strengthen the
relative importance of off-diagonal (non-Ising) terms in the Hamiltonian, which will as a result lead to coherent
quantum dynamics. The properties of such quantum spin ice will push the analyses here
into the quantum realm, which has so far only partially been
explored~\cite{Huse2003,Moessner2003b,Isakov2004,Hermele2004,Henley2005,quantum_ice,CNPF_ice2006}
and seems not unpromising.

However, even absent a breakthrough on
this front, one may hope that minor tweaking of the Hamiltonian may lead to a qualitative change in behavior of even classical spin ice systems.
Most simply, even as the relative magnitudes
of the exchange and dipolar couplings change, monopoles will start to occur more frequently in neutral bound pairs,
thereby opening a different window on the physics of two-component Coulomb liquids. Larger changes will make contact with an interesting literature on
phase transitions out of the Coulomb phase of the emergent gauge
field~\cite{Jaubert2008,3ddimertrans}.

In sum, we would contend that spin ice problem has had a fine run given the relative
simplicity of its ingredients and it appears not yet to be losing momentum!
%
%

\section*{
Disclosure statement
         }
The authors are not aware of any affiliations, memberships, funding, or financial holdings that
might be perceived as affecting the objectivity of this review.
%
%

\section*{
Acknowledgments
         }
The authors have enjoyed more discussions, with theorists and experimentalists alike, than can be
acknowledged by name. An omission of this such a list should in no way be seen as a lack of gratitude
on our behalf.  Funding was provided by NSF grant number DMR-1006608 (SLS) and
in part by EPSRC Postdoctoral Research Fellowship EP/G049394/1 (CC).
%
%


%
%




\begin{thebibliography}{99}
\bibitem{foot1}
We remind the reader that an emergent degree of freedom is one that is needed for a proper low energy description of a given system even though it has no clear meaning in its microscopic description.
\bibitem{foot2}
Although the venerable superconductor~\cite{topoSC} is historically the first example of a topologically ordered condensed matter system.
\bibitem{foot4}
We note that confinement requires that we pay an infinite energy to separate the constituents. However, deconfined objects can nonetheless exhibit bound states with finite binding energy.
\end{thebibliography}
\end{document}